\documentclass{pasa}%

\usepackage{graphicx}
\usepackage{gensymb} 
\usepackage{xfrac}   

\newcommand{\red}[1] {#1}

\title[Calibration database for the MWA ASVO]{Calibration database for the Murchison Widefield Array All-Sky Virtual Observatory}

\author[Sokolowski et al.]
{M.~Sokolowski$^{1}$\thanks{marcin.sokolowski@curtin.edu.au}, 
C.~H.~Jordan$^{2,3}$, 
G.~Sleap$^{2}$, 
A.~Williams$^{2}$, 
R.~B.~Wayth$^{1,3}$, 
M.~Walker$^{2}$, 
D.~Pallot$^{4}$, 
A.~Offringa$^{5,6}$, 
N.~Hurley-Walker$^{1}$, 
T.~M.~O.~Franzen$^{5}$, 
M.~Johnston-Hollitt$^{1}$, 
D.~L.~Kaplan$^{7}$, 
D.~Kenney$^{1}$, 
S.~J.~Tingay$^{1}$
\affil{$^1$International Centre for Radio Astronomy Research, Curtin University, Bentley, WA 6102, Australia}%
\affil{$^2$Curtin Institute of Radio Astronomy, GPO Box U1987, Perth, WA 6845, Australia}
\affil{$^3$ARC Centre of Excellence for All Sky Astrophysics in 3 Dimensions (ASTRO 3D), Australia}
\affil{$^4$International Centre for Radio Astronomy Research, University of Western Australia, Crawley 6009, Australia}
\affil{$^5$Netherlands Institute for Radio Astronomy (ASTRON), 7991 PD, Dwingeloo, The Netherlands}
\affil{$^6$Kapteyn Astronomical Institute, University of Groningen, PO Box 800, 9700 AV, Groningen, The Netherlands}
\affil{$^{7}$\red{Department of Physics, University of Wisconsin--Milwaukee, Milwaukee, WI 53201, USA}}
}%

\jid{PASA}
\doi{10.1017/pas.\the\year.xxx}
\jyear{\the\year}

\usepackage{aas_macros}
\usepackage{hyperref} 
\hypersetup{colorlinks,citecolor=blue,linkcolor=blue,urlcolor=blue}

\hypersetup{draft}

\begin{document}

\begin{frontmatter}
\maketitle

\begin{abstract}
We present a calibration component for the Murchison Widefield Array All-Sky Virtual Observatory (MWA ASVO) utilising a newly developed \textsc{PostgreSQL} database of calibration solutions. Since its inauguration in 2013, the MWA has recorded over thirty-four petabytes of data archived at the Pawsey Supercomputing Centre. According to the MWA Data Access policy, data become publicly available  eighteen months after collection. Therefore, most of the archival data are now available to the public. Access to public data was provided in 2017 via the MWA ASVO interface, which allowed researchers worldwide to download MWA uncalibrated data in standard radio astronomy data formats (CASA measurement sets or UV FITS files). The addition of the MWA ASVO calibration feature opens a new, powerful avenue for researchers without a detailed knowledge of the MWA telescope and data processing to download calibrated visibility data and create images using standard radio-astronomy software packages. In order to populate the database with calibration solutions from the last six years we developed fully automated pipelines. A near-real-time pipeline has been used to process new calibration observations as soon as they are collected and upload calibration solutions to the database, which enables monitoring of the interferometric performance of the telescope. Based on this database we present an analysis of the stability of the MWA calibration solutions over long time intervals. 
\end{abstract}

\end{frontmatter}

\section{INTRODUCTION}
\label{sec:intro}

%
The Murchison Widefield Array \citep[MWA;][]{2013PASA...30....7T,2018PASA...35...33W} is one of the low-frequency precursors of the Square Kilometer Array (SKA\footnote{https://www.skatelescope.org/}). It is located at the Murchison Radio-astronomy Observatory (MRO), the future site of the low-frequency component of the SKA (SKA-Low). The MWA began its operations in 2013 and since then has recorded over thirty-four petabytes of visibility (output from the MWA correlator) and Voltage Capture System \citep[VCS;][]{2015PASA...32....5T} data, which are archived in the Pawsey Supercomputing Centre (PSC). Based on the MWA Data Access Policy\footnote{http://www.mwatelescope.org/team/policies} the data become publicly available eighteen months after collection or immediately after collection for the members of the MWA collaboration. As a consequence, most of the archival visibility data (approximately 19.3\,PB\red{, representing 79\% of the total}) are now (as of March 2020) publicly available. Public data have been available via the MWA All-Sky Virtual Observatory (MWA ASVO\footnote{https://asvo.mwatelescope.org}) since its initial pilot release in 2017. The pilot MWA ASVO interface enabled users to download raw MWA data in standard radio astronomy data formats such as CASA measurement sets \citep{casa} or UV FITS files \citep{uvfits}. The data sets returned to the end user are flagged for radio-frequency interference (RFI) using \textsc{aoflagger} software  \citep{2012A&A...539A..95O,2015PASA...32....8O} and averaged in time and frequency according to the requirements specified in the web interface or request file. However, these data are not calibrated and require several additional steps before sky images could be formed, which could have been difficult for users not familiar with MWA data calibration. Here we present a calibration extension of the MWA ASVO, opening a new avenue for any researcher worldwide without deep knowledge of the details of the MWA instrument and data processing to download calibrated visibility data in the aforementioned formats. 

The calibration database have been populated with calibration solutions from the entire history of MWA observations. In order to do it we developed a dedicated calibration pipeline. The newly collected \red{calibration observations are automatically processed} in near real-time and the resulting calibration solutions are uploaded to the calibration database (CALDB). This enables us to monitor the stability of their phase and amplitude components, i.e. the inteferometric performance of the MWA, which allows the MWA Operations team to identify problems which may go undetected by other components of the Monitor and Control (M\&C) software\footnote{http://www.mwatelescope.org/telescope/monitor-control}. 

This paper is organised as follows. In Section~\ref{sec:calibration_database} we present the structure of the calibration database. In Section~\ref{sec:automatic_pipelines} we describe software pipelines developed to populate the database with calibration solutions derived from archived calibrator observations since 2013, and a near-real-time version of this pipeline which is used to process new calibrator observations (collected just after sunset and before sunrise). We also present a system developed for handling requests for missing calibration solutions, and web services developed to download calibration solutions for a specific observation via web browser or command line (\textsc{wget} command). In Section~\ref{sec:applications} we describe the applications of the calibration database, such as the MWA ASVO, monitoring of the interferometric performance of the MWA, providing calibration solutions to the new MWA correlator, and potential future applications for transient and ionosphere monitoring. Finally, in Section~\ref{sec:summary} we make summarising remarks and discuss the importance of these developments in the context of the future SKA-Low telescope.

\section{DATABASE OF CALIBRATION SOLUTIONS}
\label{sec:calibration_database}

\begin{figure}
	\includegraphics[width=\columnwidth]{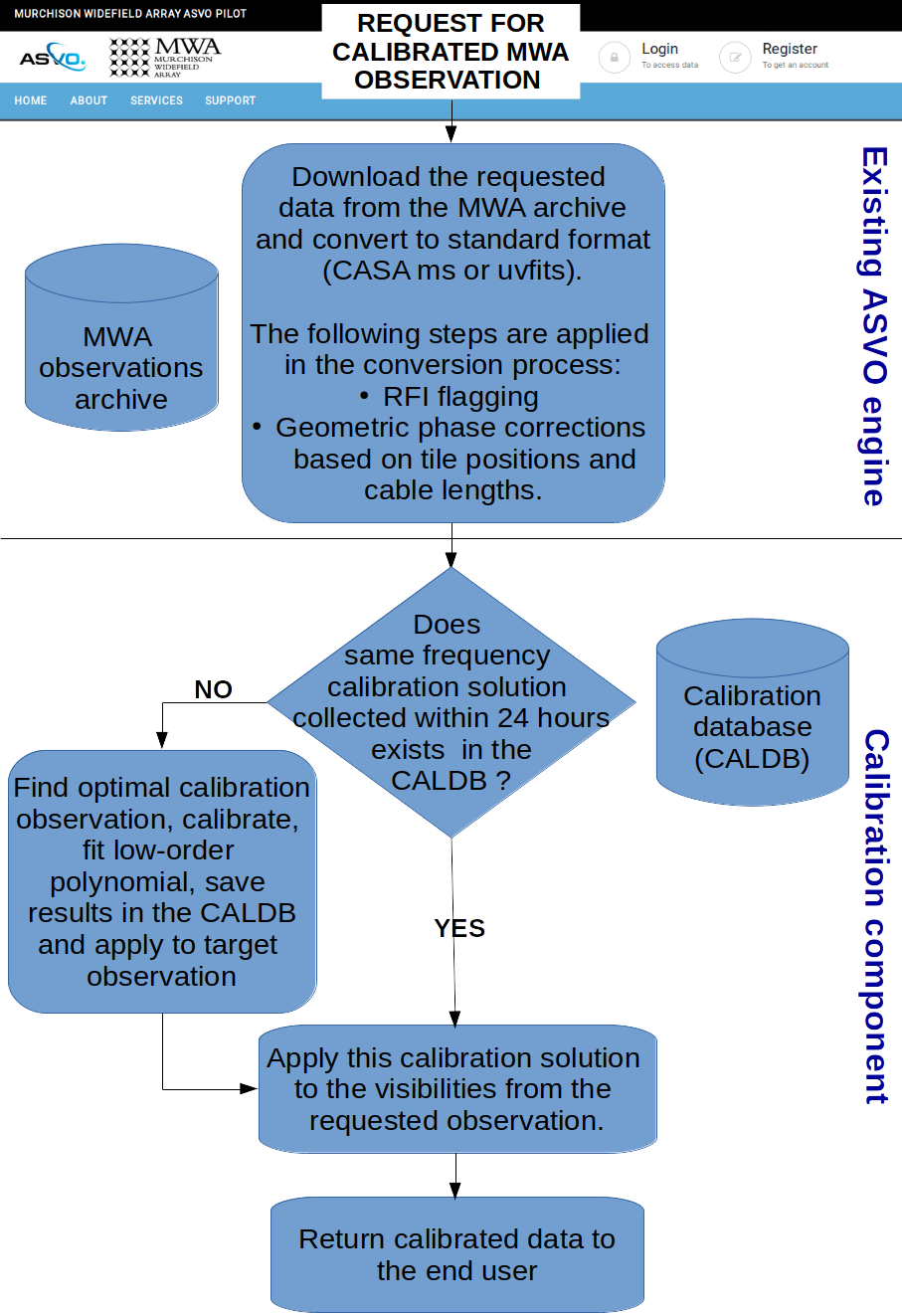}
    \caption{General overview of the MWA ASVO system with the new calibration component.}
    \label{fig_asvo_overview}
\end{figure}

The overview of the current MWA ASVO system is shown in Figure~\ref{fig_asvo_overview}. The newly added calibration component consists of the calibration database, and several scripts and pipelines for: populating this database with calibration solutions, accessing the database, and applying solutions to uncalibrated data downloaded from the MWA archive at the PSC.

\begin{figure*}
   \includegraphics[width=\textwidth]{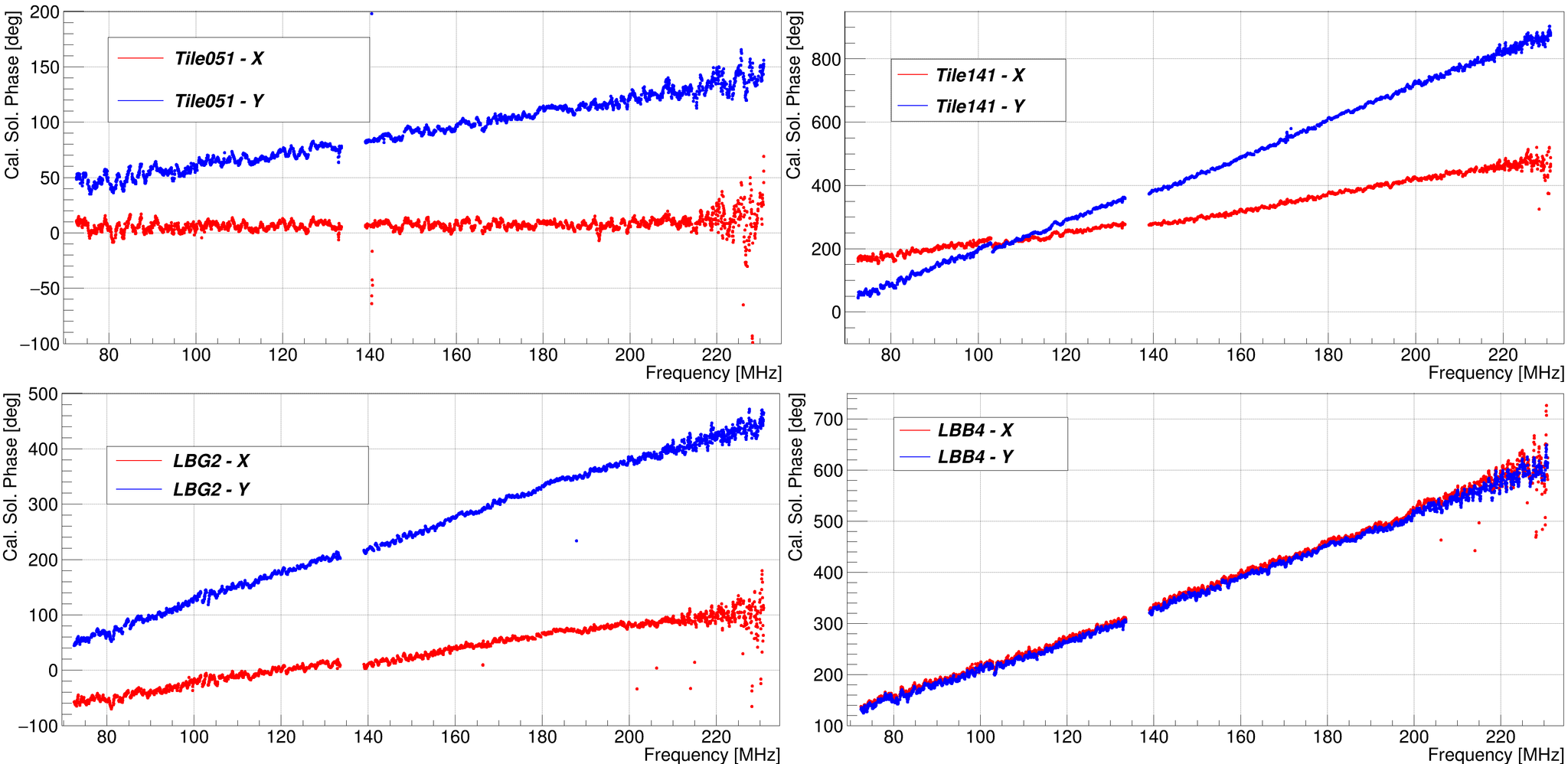}
	\caption{Calibration solution phase as a function of frequency in the range 70 - 230\,MHz for selected MWA tiles (the MWA 16-dipole units are commonly referred to as ``tiles'') in X and Y polarisations. The data were recorded on 2020-03-24 between 10:32 and 10:48 UTC. Tiles LBB2 and LBG2 are long-baselines MWA tiles. These figures show that the phase is very well modelled by a linear function of frequency over the full MWA band. The slope of a given tile depends on the electrical length of its signal path, which can be slightly different in X and Y polarisations (Tiles 051, 141 and LBG2). Frequencies above 220\,MHz are affected by radio-frequency interference.}
    \label{fig_phase_vs_freq_fullband_examples}
\end{figure*}

\subsection{DATABASE STRUCTURE}
\label{subsec:database_structure}
The calibration database (CALDB) has been implemented as a \textsc{PostgreSQL} database\footnote{https://www.postgresql.org/}, which is an advanced open source relational \red{database} and has also been used for storing other monitor and control (M\&C) MWA data. Presently the MWA can record up to 30.72\,MHz of bandwidth split into 24 coarse channels of 1.28\,MHz each. Figure~\ref{fig_phase_vs_freq_fullband_examples} shows phase of calibration solutions in the frequency range 70 - 230\,MHz computed from Pictor A observations recorded between \red{10:32 and 10:48 UTC on the 24$^{th}$ March 2020}. This figure shows that phase as a function of frequency is very well modelled by a linear function as unaccounted delays (due to cables or fibres) are the main contributors to the MWA calibration terms. Therefore, fitted parameters of a linear function provide a compact and efficient way of storing phase of calibration solutions, which is also robust against any ripples caused by reflections in cables (e.g. Tile051 in Fig.~\ref{fig_phase_vs_freq_fullband_examples} or Fig.~\ref{fig_fit_example}) or inaccuracy of the sky model used in the calibration process. This way four \red{double precision} values (two for each polarisation) are preserved for each tile\footnote{MWA 16-dipoles antenna units are commonly referred to as ``tiles''} instead of two times the number of fine channels (typically 768). Amplitude, on the other hand, can have complicated frequency structure related to the MWA tile bandpasses. However, locally (in a sufficiently narrow frequency band) it can also be approximated by a linear function; and the natural choice of the narrow band for the linear fit is the 1.28\,MHz MWA coarse channel. Therefore the database was designed to store parameters of low order (first order) polynomials fitted to amplitudes and phases of calibration solutions as a function of frequency. The original calibration solutions are stored on a hard drive and are not inserted into the database in order to keep the database compact. We envisage that this approach will likely continue in the future. The calibration database is a part of the MWA M\&C schema and consists of the following three tables (Fig.~\ref{fig_caldb_structure}):

\begin{figure*}
	\includegraphics[width=\textwidth]{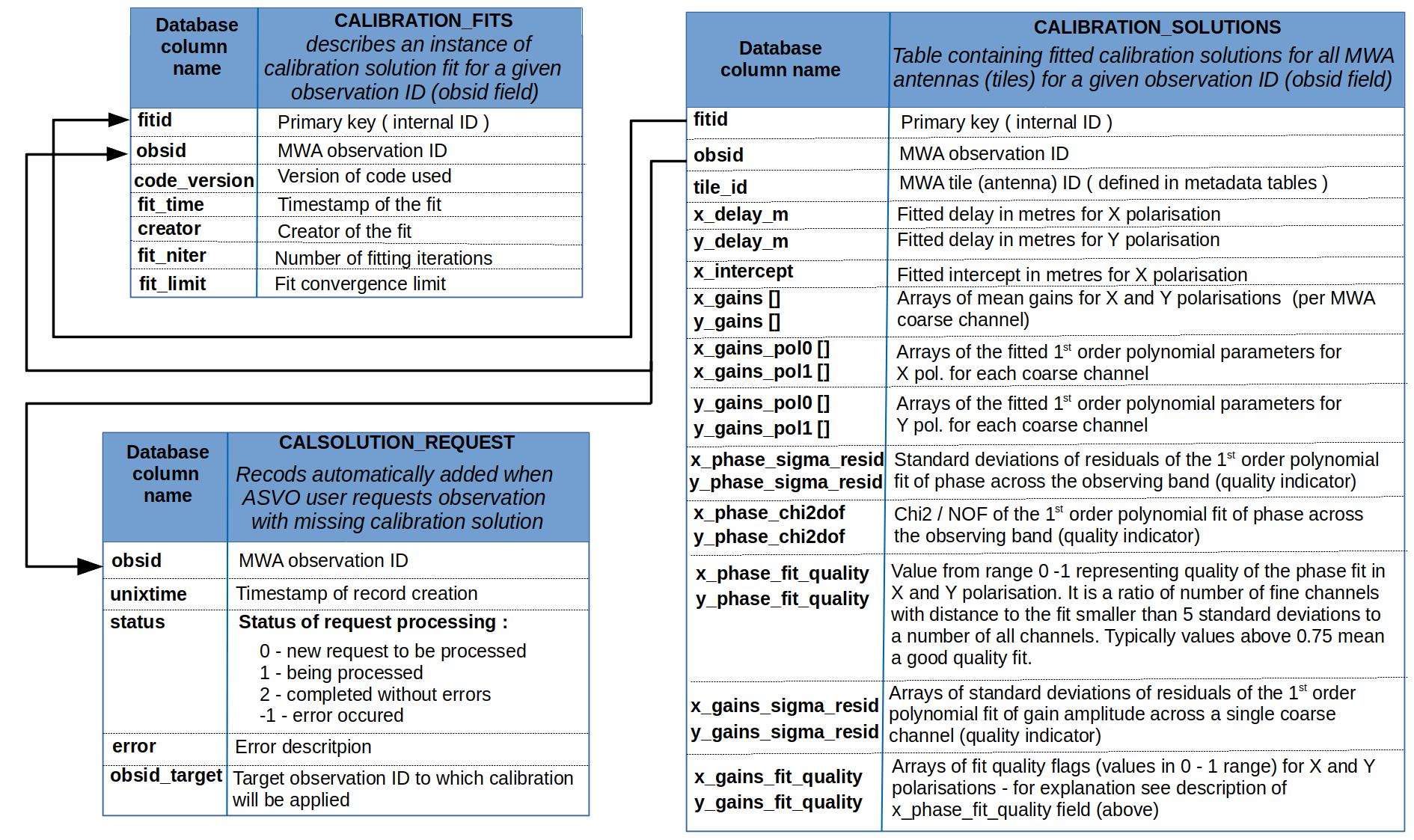}
    \caption{The tables in the MWA ASVO calibration database.}
    \label{fig_caldb_structure}
\end{figure*}

\begin{itemize}

\item \textbf{\textsc{calibration\_fits}} : provides versioning of calibration solutions stored in the database. It enables uploading newer (possibly better) calibration solutions without the necessity of removing the older versions from the database. The table contains its own unique identifier (\textsc{fitid} field), a reference to observation identifier (\textsc{obsid} field); and a timestamp of calibration solution (\textsc{fit\_time} field).

\item \textbf{\textsc{calibration\_solutions}} : each record of this table contains information about calibration solutions for both polarisations (X and Y) of a single MWA antenna. Besides references to \textsc{calibration\_fits} record (\textsc{fitid} field) and the observation identification field (\textsc{obsid}), it contains calibration fields for both polarisations with names differing by prefix in field names (x\_ or y\_). Slope (fields x\_delay\_m and y\_delay\_m) and intercept (fields x\_intercept and y\_intercept) fitted to the phase of calibration solutions are used to describe the phase of the calibration solution over the entire 30.72\,MHz of the MWA's instantaneous bandwidth, which can be either continuous or non-continuous (the latter is commonly called ``picket fence'' mode). The fitted slopes are converted to length units using speed of light in vacuum and these values are stored in the database. First-order polynomial fits (thus two database fields) are sufficient to accurately fit the phase over the full observing band provided an accurate sky model is used in the calibration process (this will be described in Section~\ref{subsec:calsols_of_archive_data}), which was verified during development and testing. Amplitudes of calibration solutions were also fitted with a linear function, but in this case the fit was performed over every 1.28\,MHz coarse channel. Therefore, they are stored as two arrays (for X and Y polarisations) of real values (typically 24, but the arrays are of variable size). This table also contains quality flags, which are real values in [0,1] range (\textsc{x\_phase\_fit\_quality} and \textsc{x\_gains\_fit\_quality} for X polarisation). These flags are calculated as ratios of the number of ``good'' channels, where the difference between the original value of a calibration solution (either phase or amplitude) and the fitted curve is smaller than 5 standard deviations ($5\sigma$), to the number of all channels. Ratio values above 0.6 are considered good quality calibration solutions. 

\item \textbf{\textsc{calsolution\_request}} : enables requests for missing calibration solutions. If the user requests calibrated data which do not have corresponding calibration solutions for the same frequency channels within 12\,hours in the calibration database, a new request record is inserted into the table \textsc{calsolution\_request} (if it is not already present there). Then an automatic script finds all the new records in this table, identifies corresponding calibration observations in the main MWA database, calibrates them, and uploads resulting calibration solutions into the calibration database. If the appropriate calibration observations cannot be found or the calibration procedure fails, the error message is stored in the field \textsc{error} and can be returned to the end user. 
\end{itemize}

\subsection{PRESENT STATUS OF THE DATABASE}
\label{subsec:status_of_the_db}

Currently (as of March 2020) the database contains calibration solutions from around 11,200 calibration observations, which provides, on average, five calibration solutions per day; one for each of the primary MWA frequency bands (at centre frequencies of 88.32, 119.04, 154.88, 185.6, and 216.32\,MHz). The database grows every day as new calibrator observations are collected and the near-real time pipeline calibrates them and inserts calibration solutions into the database.

In order to populate the database with the historic and new calibration solutions, we developed a dedicated reduction pipeline, described in Section~\ref{sec:automatic_pipelines}. 


\section{AUTOMATIC CALIBRATION PIPELINES}
\label{sec:automatic_pipelines}


Originally the pipeline used \textsc{CASA} software to calibrate calibrator observations in near-real time and create control images of the calibrator observations. In order to calibrate many archival calibrator observations, we developed a new pipeline using software more suited to the MWA observations. We are upgrading the current contents of the database with calibration solutions from the new pipeline in order to create a uniform database of calibration solutions resulting from the same data reduction pipeline, software and sky model. Both pipelines use the MWA ASVO interface to download uncalibrated \textsc{CASA} measurement sets of calibrator observations, which are produced on the MWA ASVO severs. As described in Section~\ref{subsec:applying_solutions}, the \textsc{cotter} program is used in the conversion process, which implies that RFI flagging is also applied at this stage.

\subsection{Near real-time calibration of new calibration observations}
\label{subsec:real_time_calibration}

The CASA-based pipeline has been used to reduce newly collected MWA calibrator observations. Every day the MWA observes a calibrator source shortly after sunset and before sunrise at five standard frequency bands (at centre frequencies of 88.32, 119.04, 154.88, 185.6, and 216.32\,MHz) and in the so-called ``picket fence'' mode with 24 coarse channels spread regularly over the frequency range $78 - 240$\,MHz. The calibrator script continuously runs on one of the MWA servers, checks the MWA schedule database for new calibrator observations and whenever it detects that new calibrator observations were collected they are automatically downloaded, calibrated and control images of the calibrator field are formed at selected frequencies. Presently the observations are downloaded from \red{the PSC and processed} on a server at Curtin Institute of Radio Astronomy (CIRA), which introduces additional delay. In the future, this processing will be relocated to the MRO, and the pipeline will use ``raw'' visibility files as they are produced by the MWA correlator. If the quality of the resulting calibration solutions satisfies minimum requirements they are uploaded to the calibration database. \red{The requirements} for the new calibration solutions to be loaded to the database are the following: (i) they must be better than the ones already in the database (if there are any for this \textsc{obsid}) and (ii) more than half of antennas have acceptable calibration solutions, where ``acceptable'' means that more than 60\% of the channels have a phase fit within $5\sigma$ from the data (this criteria is to avoid storing calibration solutions of very low quality). These near-real time calibration solutions are used for monitoring of the interferometric performance and enable us to examine the long term stability of the MWA (Sec.~\ref{sec:applications}).

\subsection{Calibration solutions of archived data}
\label{subsec:calsols_of_archive_data}

The calibration database has been populated with calibration solutions starting from the beginning of 2013. In order to achieve this we created a list of all calibrator observations and submitted them for processing by the calibration reduction pipeline \textsc{Heracles}\footnote{\url{https://gitlab.com/chjordan/heracles}}. The pipeline is using \textsc{calibrate} software \citep{offringa-2016} upgraded with the newest 2016 MWA beam model \citep{beam2016} and sky model generated by PUMA\footnote{\url{https://github.com/JLBLine/srclists}} \citep{puma2018}. It creates binary files with calibration solutions and control images of the field using the \textsc{WSCLEAN}\footnote{\url{https://sourceforge.net/p/wsclean/wiki/Home/}} program \citep{OffMcK14}.
 
\subsubsection{HERACLES pipeline}
\label{subsec:heracles}

Our initial attempt for calibration of the MWA archive used a single virtual machine on the cloud-based system ``Nimbus'' hosted by the PSC. However, our task quickly proved to be insufficient for the amount computing resources afforded by this system. For this reason, \textsc{heracles} was converted into a generalised and distributed system, which could be run on any free resources available (such as unused desktop computers in CIRA).

\textsc{heracles} is primarily utilised by a single executable which has two
modes of operation: server and client. The server mode is primarily concerned
with which observations need to be calibrated, based on a \textsc{SQLite}
database. This database tracks which observations have not yet been calibrated,
which have been calibrated, and which have failed. The \textsc{heracles} server
must also coordinate with the state of any MWA ASVO data downloads (available,
in progress, not available, etc.). So as to not flood the MWA ASVO with download
requests, the \textsc{heracles} server uses a run-time setting to prepare a
certain number of observations for download as the clients progress.

Once connected to a server, the operation of a \textsc{heracles} client follows
a simple loop:
\begin{enumerate}[label= (\roman*),leftmargin=\parindent]
\item Request an observation to calibrate. If an observation is ready, move to
  step (ii), otherwise, wait for one minute before querying the server again;
  \item Download the observation;
  \item Operate upon the data with an executable (set at run-time, typically a
    \textsc{bash} script);
  \item If the result of the executable was a success, the calibration solutions
    and any other useful products are transmitted to the server. Return to step
    (i); and 
  \item If any failure occurs in the loop, it is also reported to the server,
    before returning to step (i)
\end{enumerate}

As the computational load of the server is negligible, clients may also be run
on the same computer as the server.

\begin{figure*}
    %
    %
    %
	\includegraphics[width=\columnwidth]{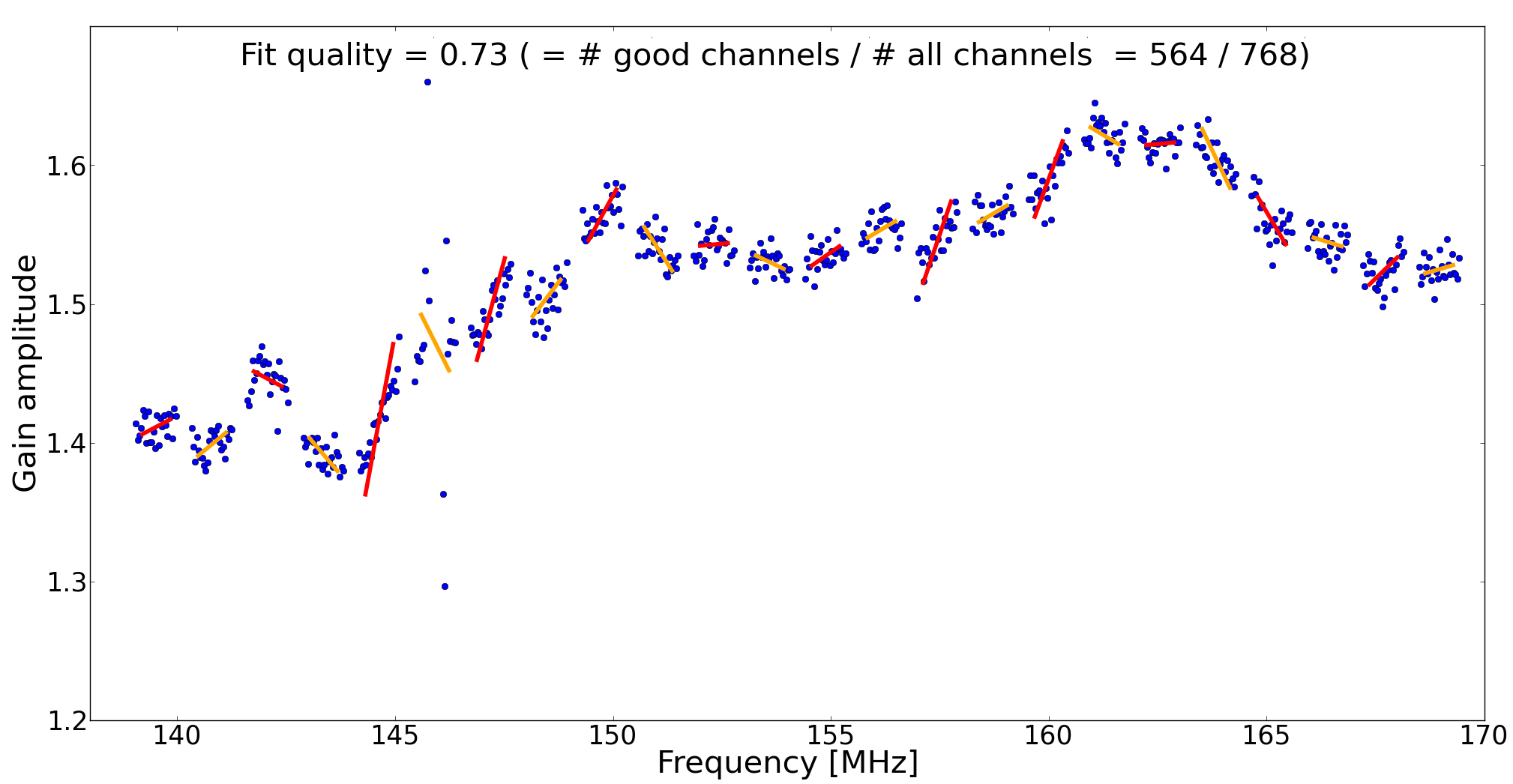}
	\includegraphics[width=\columnwidth]{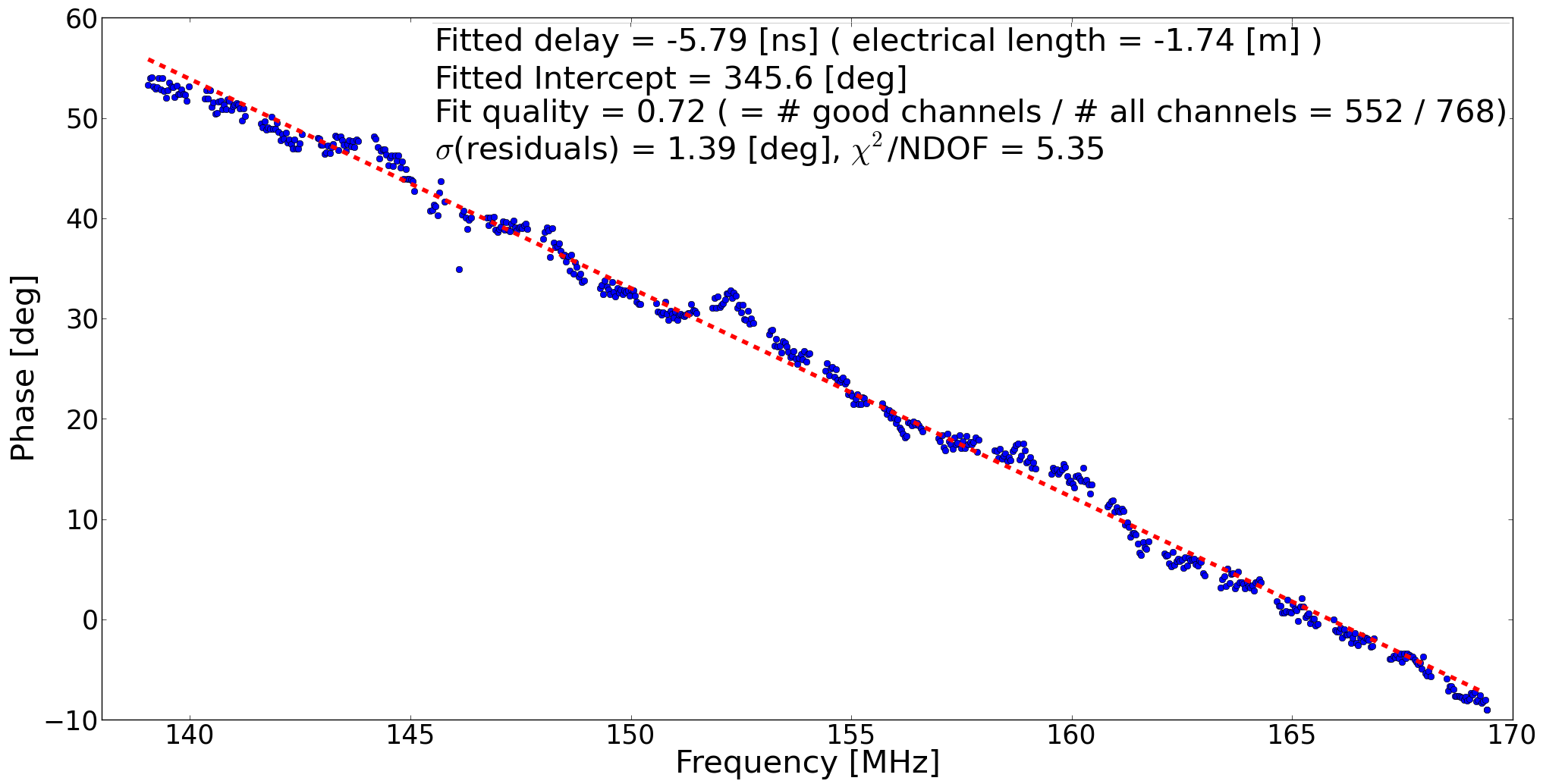}
	\includegraphics[width=\columnwidth]{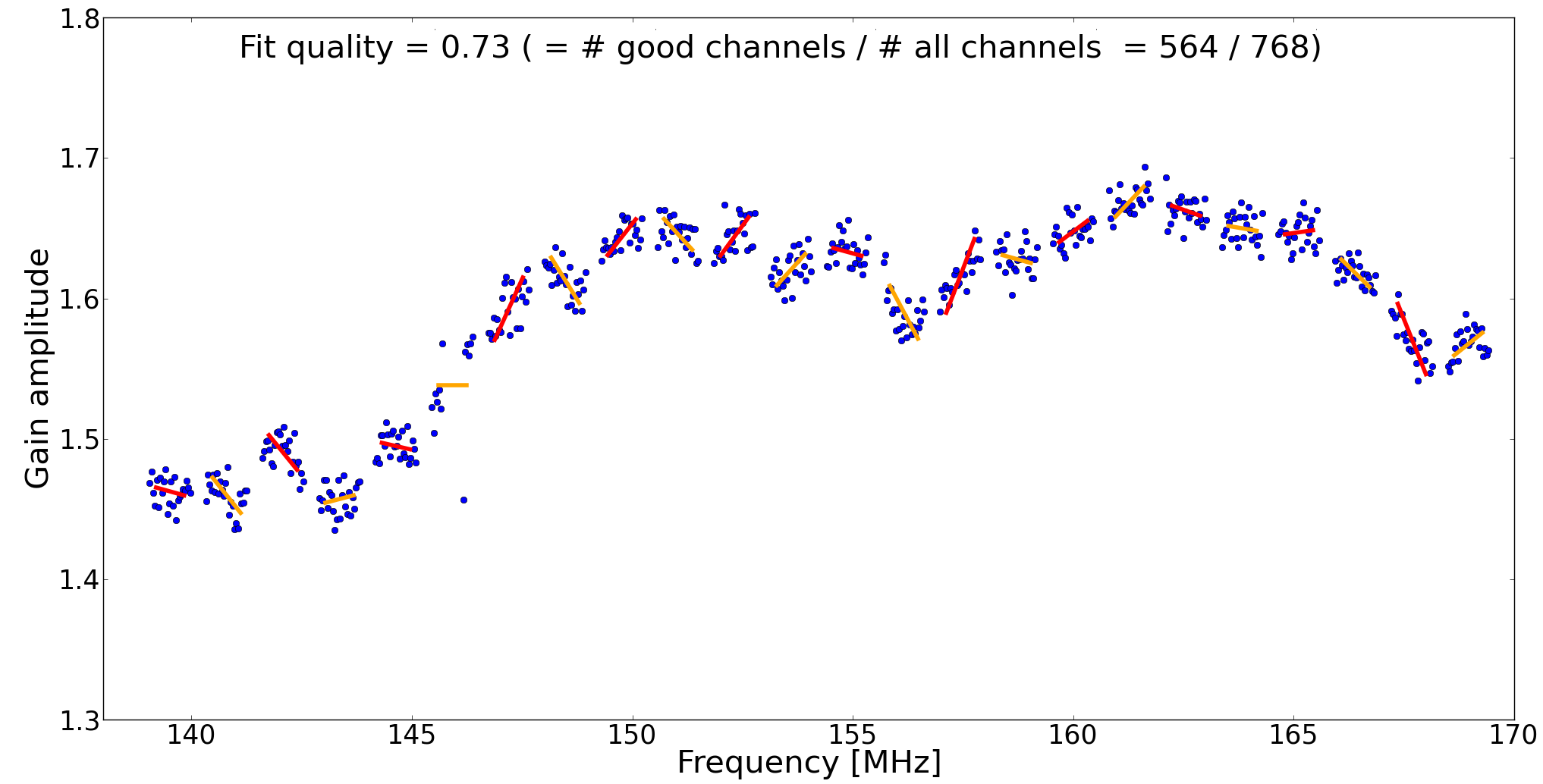}
	\includegraphics[width=\columnwidth]{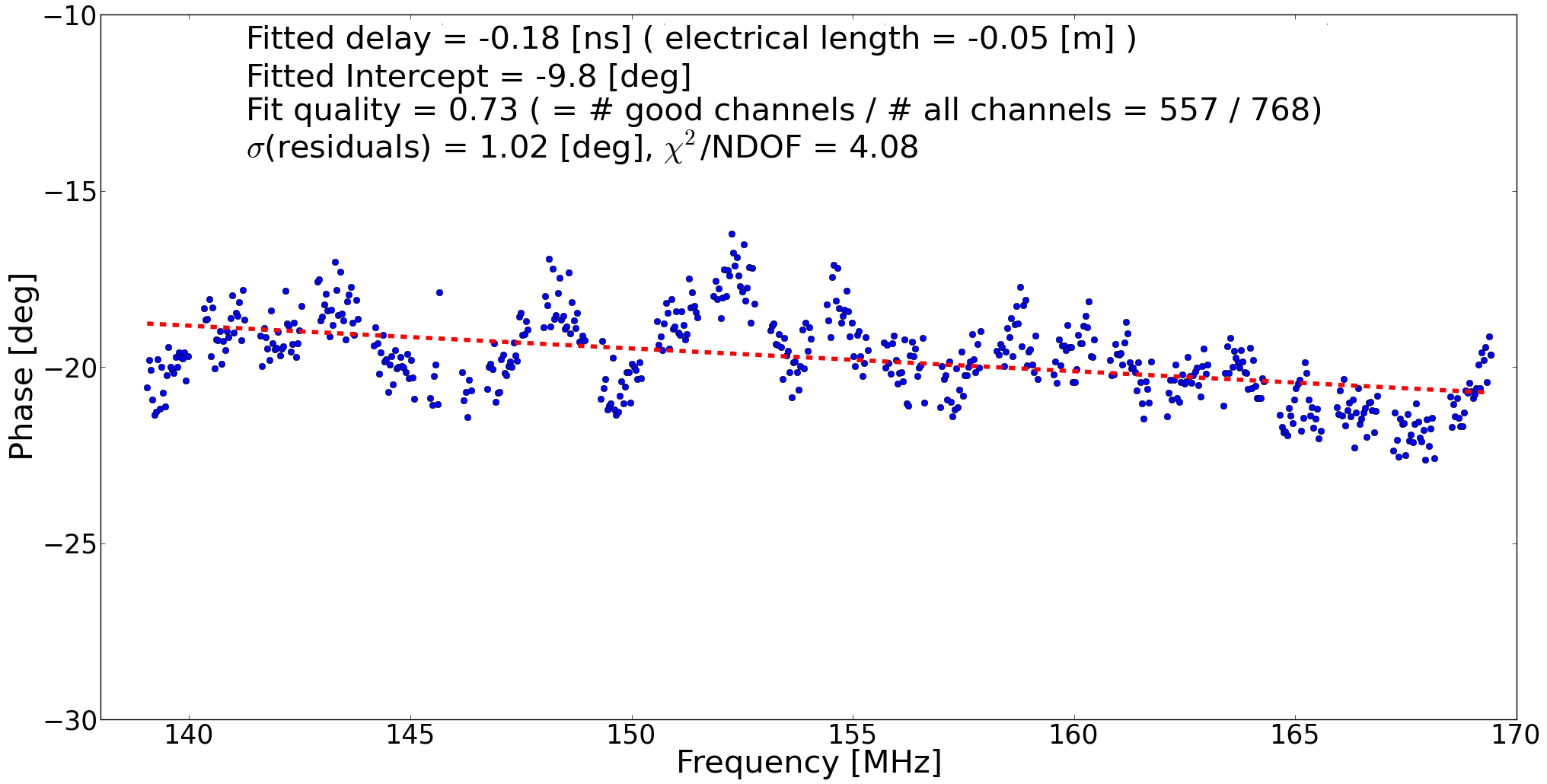}
    \caption{Examples of calibration solutions fitted with first order polynomials for MWA tiles 136 (upper row) and 065 (lower row). These type of plots are automatically generated whenever new set of calibrator observations are collected (just before sunrise and shortly after sunset). Left panels show examples of first order polynomials fitted to the gain amplitude in each of the 24 MWA coarse channels. Right panels show examples of linear functions  fitted to phase over the MWA continuous observing band of 30.72\,MHz. It can be noted that amplitude fitting may not work well for coarse channels affected by radio-frequency interference (RFI), and a combination of wider band fitting with more fit/exclude outliers iterations can further improve the final quality of the fitted amplitudes of calibration solutions.}
    \label{fig_fit_example}
\end{figure*}

The mode of operation of \textsc{heracles} clients allows users to enable or
disable clients dynamically, which allows otherwise unused computing resources
to be utilised, and proved to be an effective, efficient solution for
calibrating a large volume of data. Within a few months, we were able to
download, calibrate, and image observations and insert solutions into the
calibration database from nearly six years of MWA operations.

\subsubsection{CASA pipeline}
\label{subsec:casa}

Originally, the \textsc{CASA}-based pipeline was used to calibrate evening and morning calibration scans and store the calibration solutions in the database. This pipeline used VLA Low-Frequency Sky Survey Redux (VLSSr; \citet{2014MNRAS.440..327L}) images of calibrator sources (such as Hydra A, 3C444, Hercules A and Pictor A) to derive calibration solutions. Since the creation of the new \textsc{heracles} pipeline, the \textsc{CASA} pipeline will be retired and the calibration solutions in the database are being superseded with the results from the new pipeline.

\subsection{Uploading calibration solutions to the database}
\label{subsec:uploading}

Phase and amplitude of calibration solutions resulting from the reduction pipeline are fitted with the first order polynomial as a function of frequency. 
Figure~\ref{fig_phase_vs_freq_fullband_examples} shows that phase of calibration solutions is a linear function of frequency over a very wide band. The lowest and highest four 40\,kHz fine channels (160\,kHz) in each coarse \red{channel} as well as the fine channels flagged due to RFI (during the conversion process) are excluded from the fitting.

First, the phase of calibration solutions over the 30.72\,MHz band is ``unwrapped''; phase values are not limited to $[-180,+180]$ degrees, but can range from minus infinity to plus infinity. Then the phase is fitted with a linear function over the entire observing band (30.72\,MHz in continuous observations) resulting in two fit parameters: slope and intercept (right column in Fig.~\ref{fig_fit_example}). These are sufficient to accurately describe the phase of calibration solutions as a function of frequency, provided that the sky model used in the calibration process is complete (this was verified in the development and testing stage). The fitted slope is converted to a corresponding time delay ($\Delta t$) and eventually the length $c \Delta t$ (in metres), where $c$ is speed of light in vacuum, is saved to the calibration database. 

The amplitude of calibration solutions is also fitted with a first order polynomial and in this case the fit is performed over a single 1.28\,MHz coarse channel, resulting in different slopes and intercepts for each of the 24 coarse channels (left column in Fig.~\ref{fig_fit_example}). It was verified that linear fit is the optimal polynomial order to fit amplitudes over an MWA coarse channel as a parabola had only slightly lower $\chi^2$ values and nearly two times higher Bayesian Information Criterion value (BIC; \citet{1978AnSta...6..461S}), which proved that the linear fit is a more appropriate representation of data than the parabola.

\begin{figure}
    \includegraphics[width=\columnwidth]{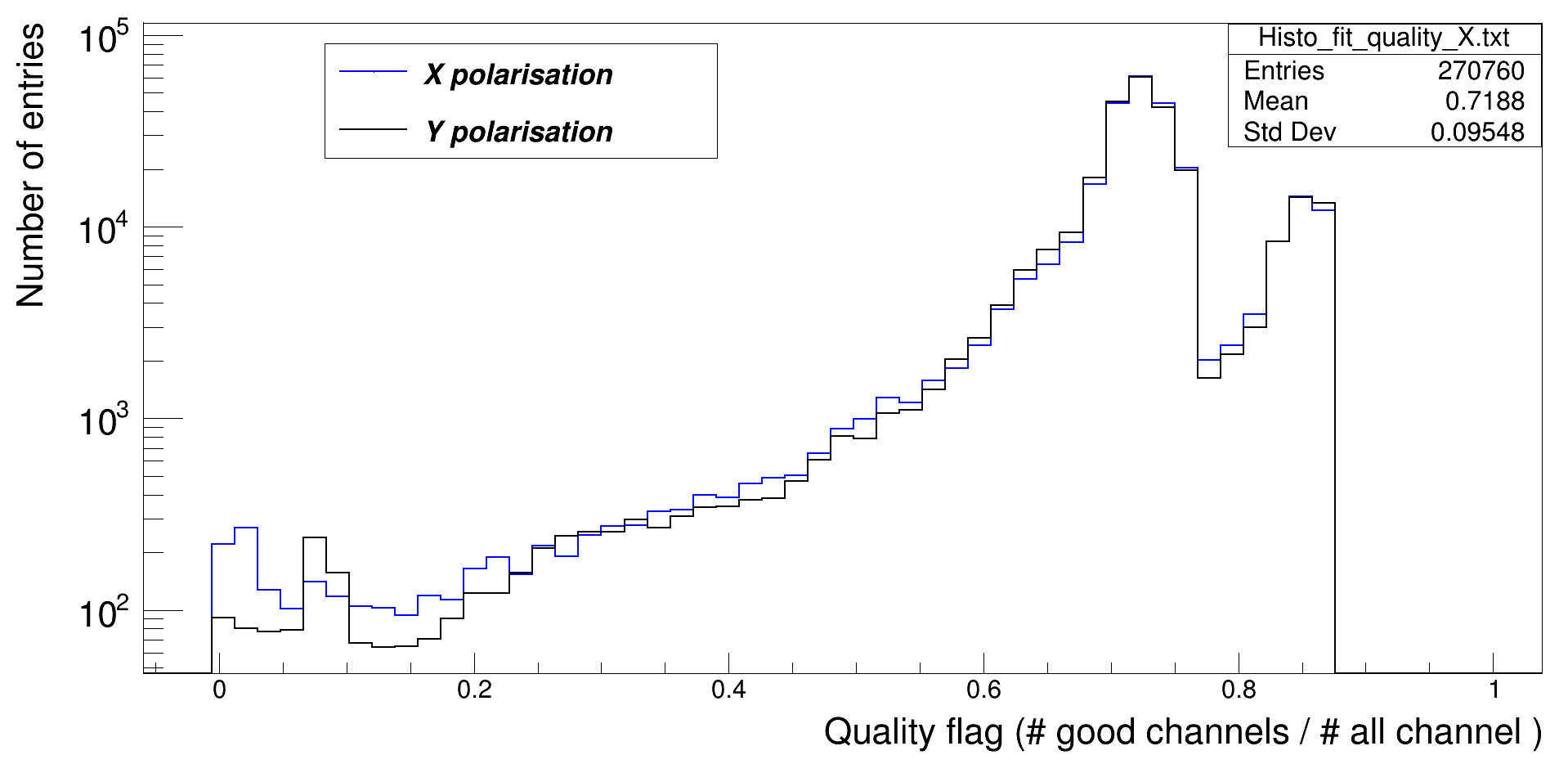}
    \caption{Distribution of quality flag for calibration solutions in the MWA ASVO database for X and Y polarisations. Quality flag does not exceed $\approx$0.88 because fine channels at the edges of coarse channels are always flagged. The dual peak structure is due to the fact that calibration solutions in coarse channel 121 (154.88\,MHz), which is the optimal MWA operating frequency, are noticeably better than at other frequencies and they generate a peak at around 0.85. There are 419 observations with flag quality exceeding 0.75 for all antennas (in both polarisations) and these are observations of : 3C444 (224), Pictor A (59), Centaurus A (45), Hydra A (29), PKS2356-61 (27) and a drift scan field (11), where the numbers in brackets are the number of observations for the particular calibrator. This indicates that majority of best quality calibration solutions comes from observations of 3C444. The main peak at around 0.73 is due to good quality calibration solutions at other standard frequencies (88.32, 119.04, 185.6 and 216.32\,MHz).}
    \label{fig_quality_distribtion}
\end{figure}

If the fit satisfies quality requirements the resulting fitted parameters are stored in the database. The current quality requirement is that the ratio between number of good quality channels to all the channels in the calibrated observation is above 0.6 (Figure~\ref{fig_quality_distribtion}).


\subsection{Accessing calibration solutions in the database}
\label{subsec:getting_solutions}

The calibration solutions in the database can be accessed via a web service with a standard \textsc{wget} command\footnote{wget http://mro.mwa128t.org/calib/get\_calfile\_for\_obsid?
obs\_id=OBSID\&zipfile=1\&add\_request=1 where \textsc{OBSID} should be replaced with the observation ID of the target or explicit calibrator observation.}.
The request is executed on an MWA server and if the appropriate (the same frequency band and within 24\,hours from the target observation) calibration solution exists in the database it is returned to the user in the same binary file format (``.bin'' file) as produced by the calibration procedure developed by \citet{2016MNRAS.458.1057O}. If there is no suitable calibration solution an error message stored in a text file is returned to the end user.

\subsection{Application of calibration solutions to data downloaded from the MWA ASVO}
\label{subsec:applying_solutions}
The MWA ASVO website and API\footnote{https://github.com/ICRAR/manta-ray-client} allow users to submit ``conversion'' jobs which, when run, retrieve the observation, pre-process the data, converting the raw MWA correlator visibility format into a standard CASA or UV fits format, and then make the data product available for download. The conversion/pre-processing steps expose the options available in the \textsc{cotter} pre-processing pipeline \citep{2015PASA...32....8O}. The MWA ASVO calibration option utilises a recently-added feature of \textsc{cotter}, allowing calibration solutions retrieved from the calibration database to be applied to the data before any data averaging takes place.

In a typical conversion job with the calibration option set, the requested observation is staged from the Pawsey Long Term Archive (LTA). The LTA has a Hierarchical Storage Management (HSM) system consisting of several different tiers of storage, ranging from 1.5 PB of spinning disk to an allocation of 40 PB of magnetic tape. Once the observation data are available on the disk cache they are copied to a scratch area. A web service call is made to retrieve the metafits file that contains much of the metadata associated with the requested observation - this is also stored with the observation files in the scratch area.

The calibration web service is called by the MWA ASVO to retrieve the best calibration solution for this observation (Sec. \ref{subsec:getting_solutions}). If the solution is found, then the calibration solution binary file is retrieved and stored with the rest of the observation files in the scratch area. If no calibration solution is suitable, then the job fails and a request record is added to the calibration database to produce a solution for this observation. The user is informed to try again once this is complete (usually within 24 - 48 hours).

With all of the files now available, the server then executes \textsc{cotter}, with the ``--full-apply'' command line argument, which applies the provided calibration solution to every integration before averaging (if requested). There is also another new \textsc{cotter} option ``--apply'' which applies provided calibration solutions after averaging integrations over a requested interval. Once \textsc{cotter} has produced the output data in a standard radio-astronomy data format, a download \textsc{url} is provided to the user via the website or API so the data can be retrieved. During the conversion process RFI flags (either pre-computed by \textsc{aoflagger} or calculated by \textsc{cotter}) are also applied to the data. Hence, the resulting data sets do not require any further pre-processing and initial sky images can be formed using standard radio astronomy software tools, such as for example \textsc{WSCLEAN}, \textsc{CASA} or \textsc{MIRIAD} \citep{1995ASPC...77..433S}. These initial images can be used in self-calibration procedure in order to improve calibration solutions and/or further processing steps, such as primary beam\footnote{https://github.com/MWATelescope/mwa\_pb} or ionospheric corrections\footnote{https://github.com/nhurleywalker/fits\_warp}, may be applied depending on the requirements of the specific science case.

\section{OTHER APPLICATIONS OF THE CALIBRATION PIPELINES AND DATABASE}
\label{sec:applications}

Besides the main application of the CALDB, which is to enable downloading of calibrated data in standard astronomy data formats via the MWA ASVO interface, there are several other benefits of having a complete database of calibration solutions, which will be described in this section.

\subsection{Monitoring performance and stability of the MWA telescope}
\label{subsec:monitoring_performance}

%
%
%
\begin{figure*}
    \includegraphics[width=\textwidth]{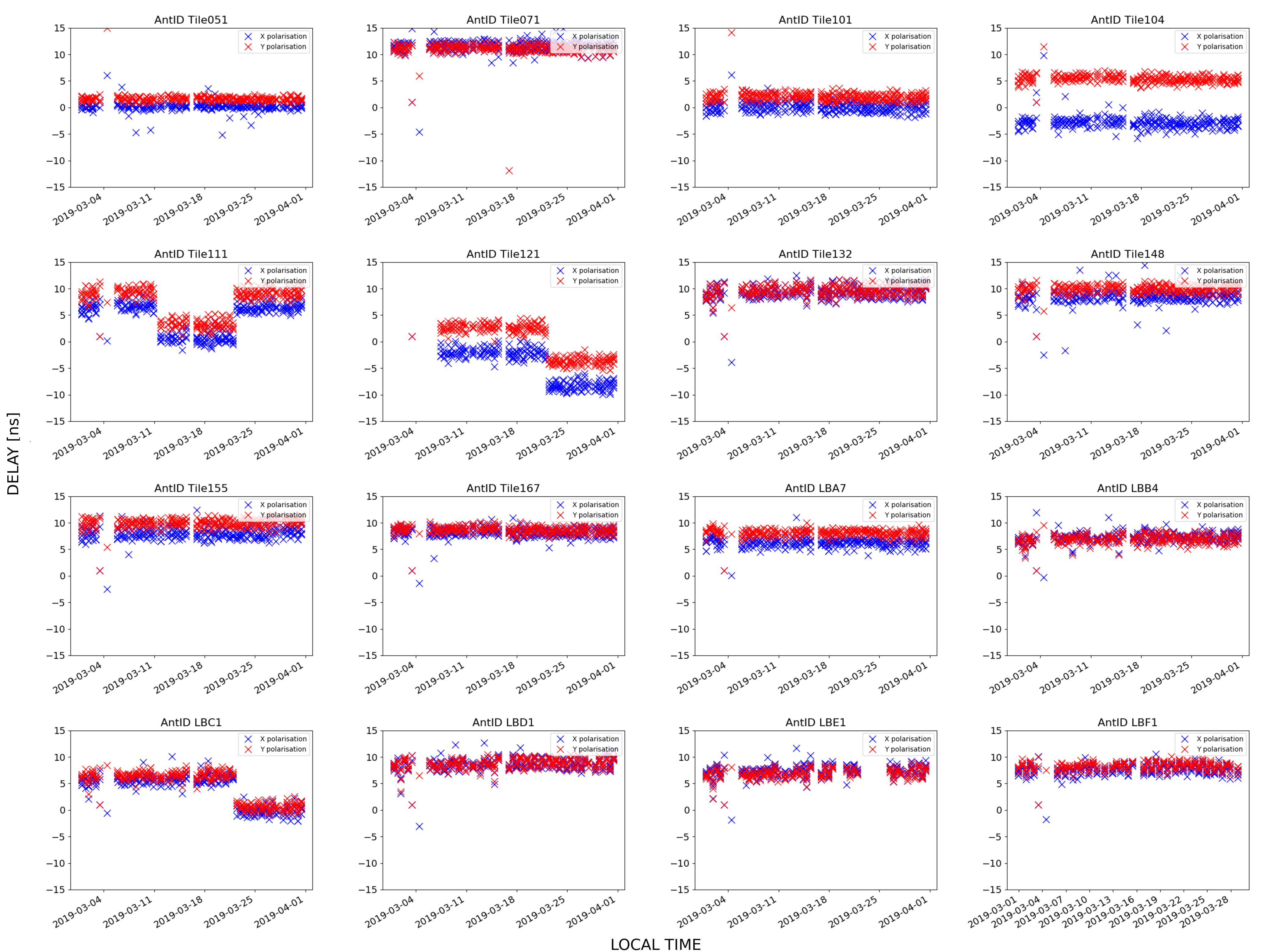}
    \caption{Delay (ns) fitted to the phase of calibration solutions as a function of time, in March 2019 when the MWA was in the extended configuration \citep{2018PASA...35...33W}. Sixteen tiles representing typically observed behaviour are shown. The sudden ``jumps'' by approximately 6\,ns (for tiles 111, 121 and LBC1) were due to receiver resets (affects all eight tiles connected to a particular receiver). The delay usually returns to the pre-reset value after the next receiver reboot. Because of this effect, the cable delays are usually not adjusted to an accuracy better than 4\,m (1\,m corresponds to time delay of $\approx$3.33\,ns). It can be seen that when receivers are not rebooted the telescope remains very stable over timescales of weeks.}
    \label{fig_mwa_month_stability_extended}
\end{figure*}

\begin{figure*}
   	\includegraphics[width=\columnwidth]{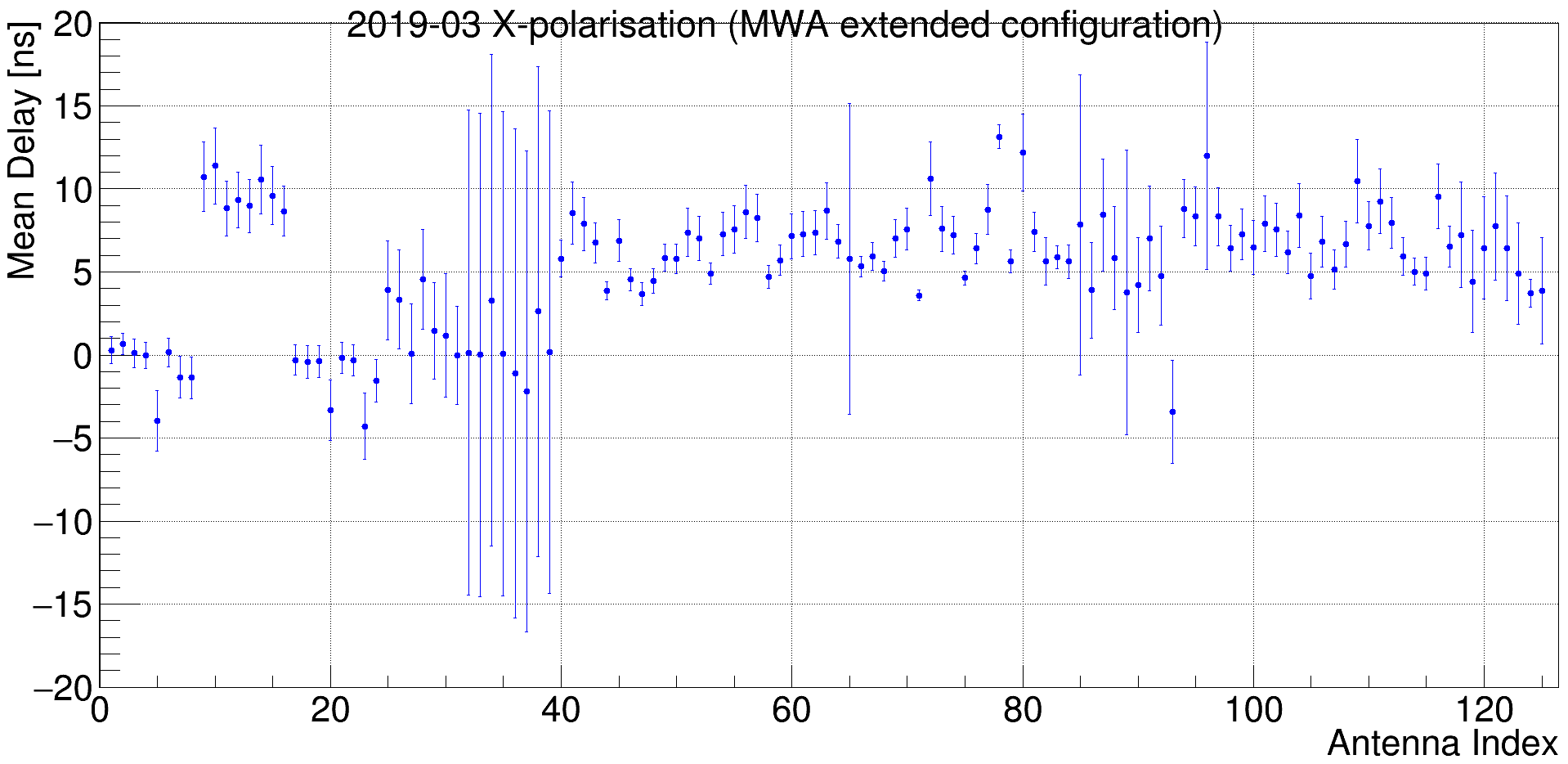} 
   	\includegraphics[width=\columnwidth]{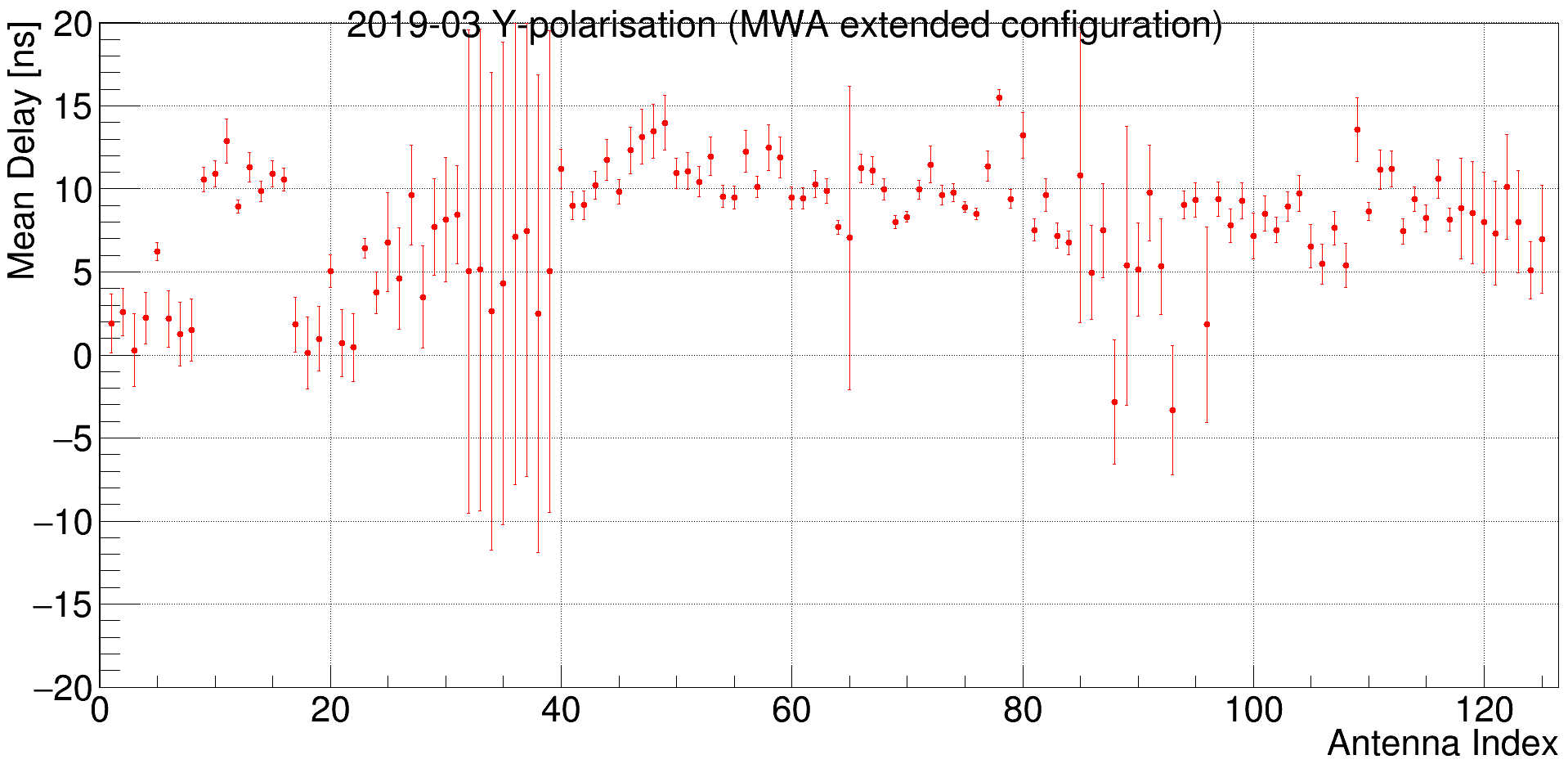} 
    \caption{Mean and standard deviation (error bars) of delays for each MWA tile during March 2019 (the MWA in extended configuration). Left image (blue points) is X-polarisation and right image (red points) is Y-polarisation. These plots are based on data points as shown in Figure~\ref{fig_mwa_month_stability_extended} for all MWA tiles and are good summary diagnostic plots of the telescope stability. For instance, indexes 9-16 (values around 10\,ns) from Tiles 071 to 078 had unaccounted cable length of around 3.23\,m, which was not corrected over that month. Large errors in indexes from 32 - 40 (Tiles 121 - 128) are due to step-like receiver reboots during the month and the same applies to indexes 25 - 32 (Tiles 111 - 118). Indexes 65 and 89 (Tile62 and LBC4 respectively) had cable lengths adjusted in March 2019 corresponding to 18.5 and 5\,ns (thus large error bars).}
    \label{fig_mean_and_stddev_201903_xy}
\end{figure*}

%
%
%
\begin{figure*}
    \includegraphics[width=\textwidth]{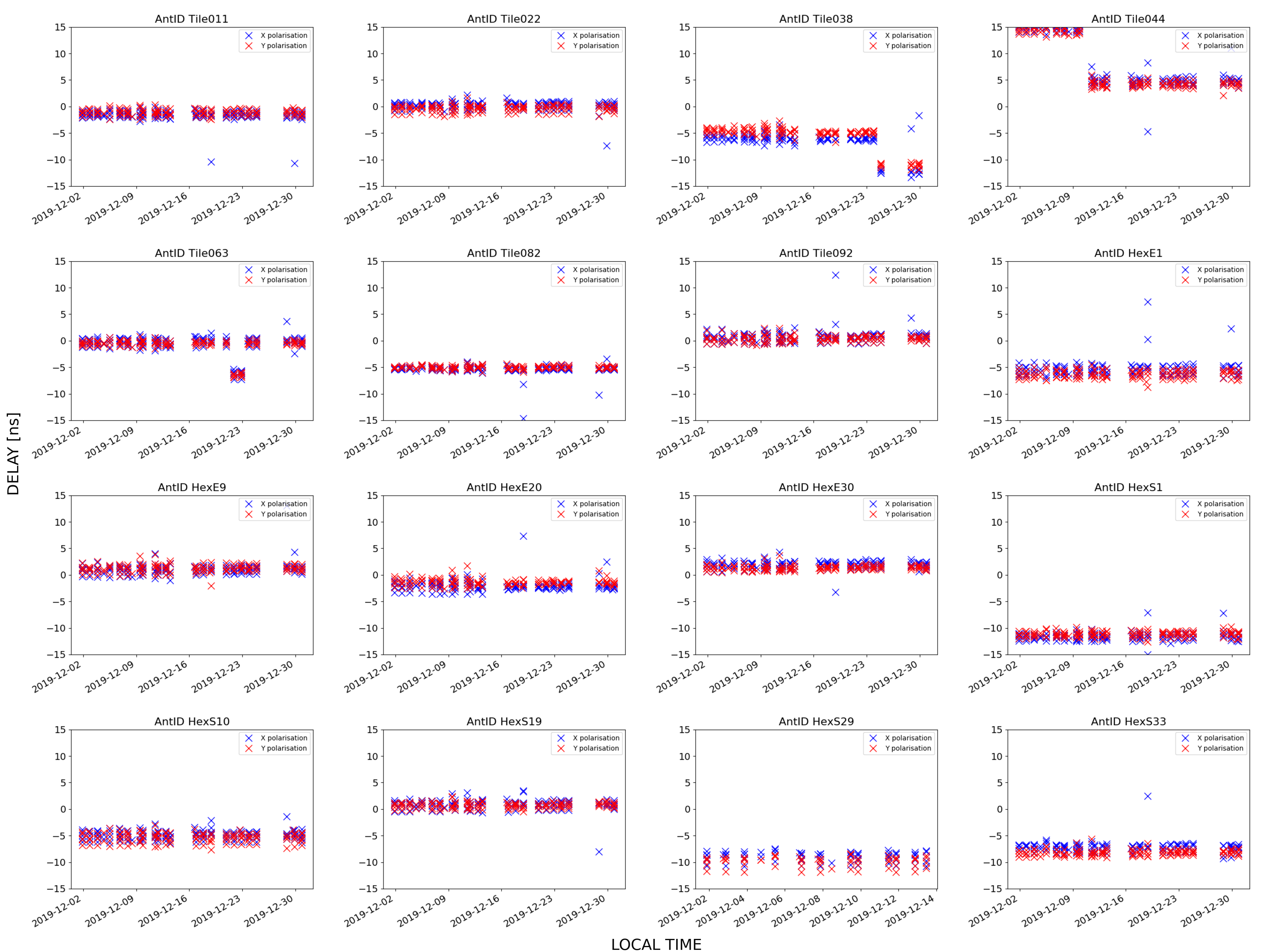}
    \caption{Delay (ns) fitted to the phase of calibration solutions as a function of time in December 2019 (with the MWA in the compact configuration) \citep{2018PASA...35...33W}. Sixteen tiles representing typically observed behaviour are shown. The short sudden "dip" of about 6\,ns observed on Tile063 was due to a receiver reset (affects all eight tiles connected to a particular receiver). The delay returned to the pre-reset value after a few days when the next reboot occurred. A step of about 10\,ns on Tile044 was due to the adjustment of its cable length in the database. It can be seen that when receivers are not rebooted the telescope remains very stable over timescales of weeks.}
    \label{fig_mwa_month_stability_compact}
\end{figure*}

\begin{figure*}
   	\includegraphics[width=\columnwidth]{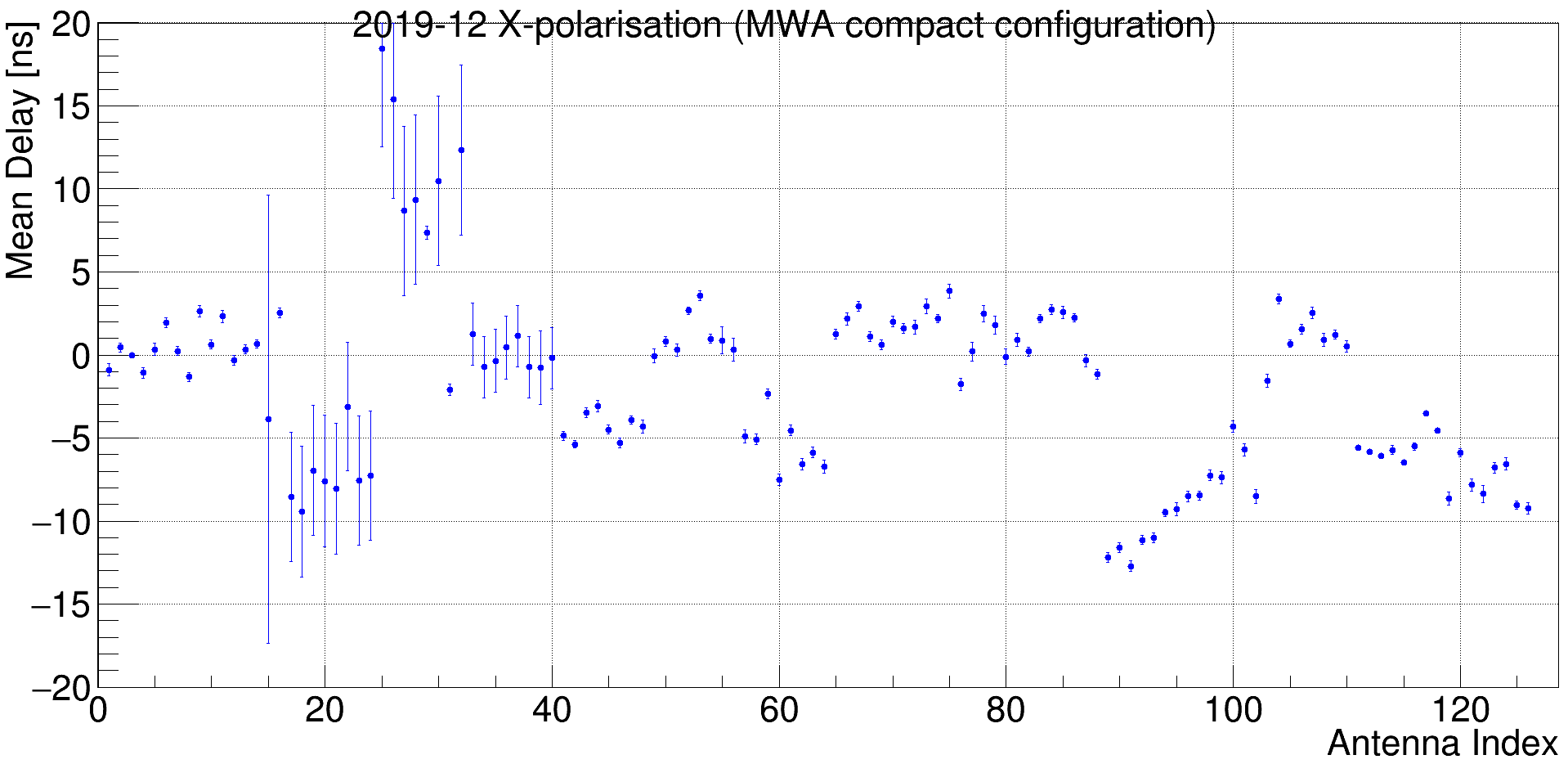} 
   	\includegraphics[width=\columnwidth]{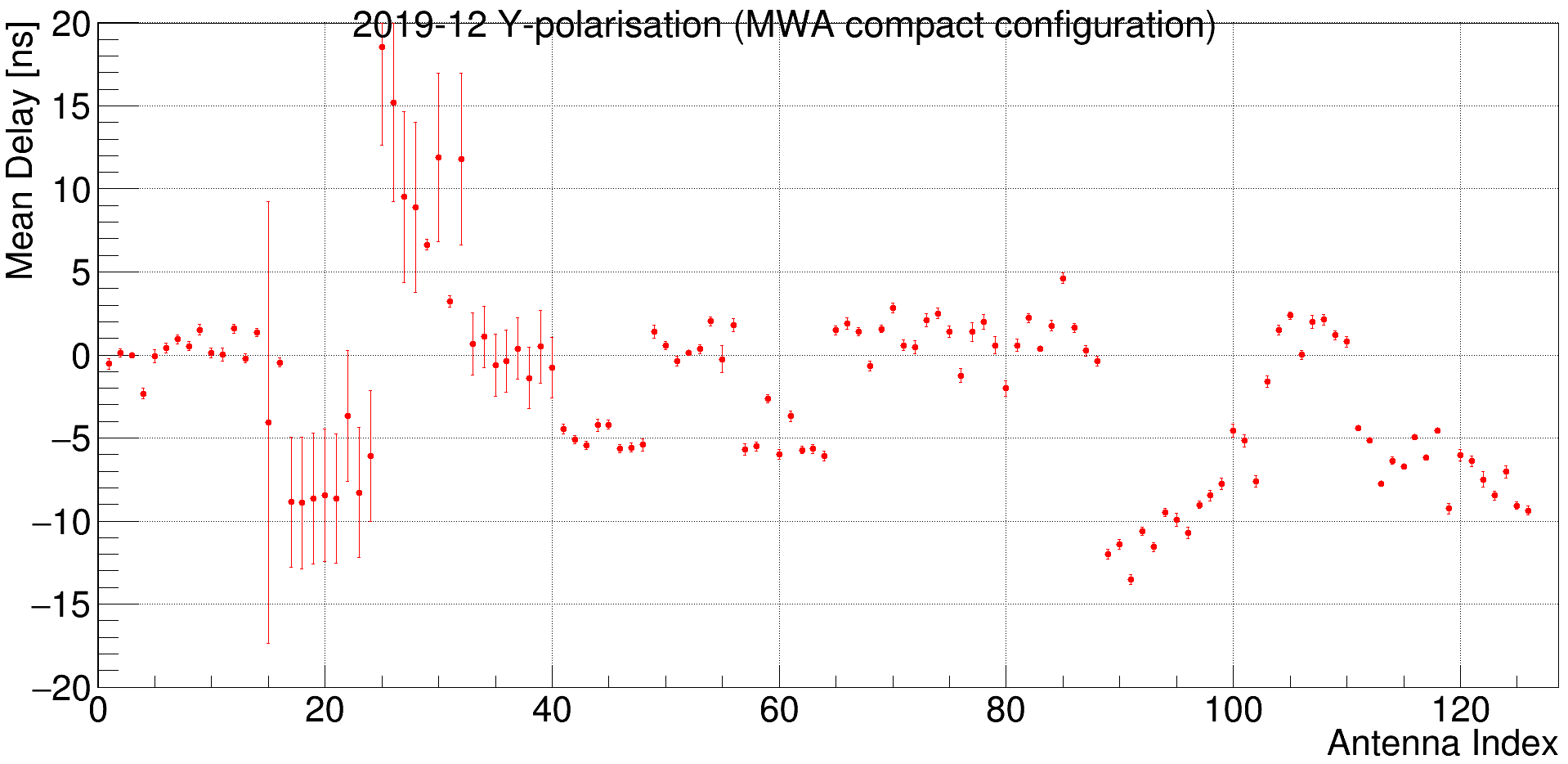} 
    \caption{Mean and standard deviation (error bar) of delays for each MWA tile during December 2019 (the MWA in compact configuration). Left image (blue points) is X-polarisation and right image (red points) is Y-polarisation. These plots are based on data points shown in Figure~\ref{fig_mwa_month_stability_compact} for all MWA tiles and are good summary diagnostic plots of the telescope stability. For instance, indexes 17 - 24 (Tiles 031 - 038) and 25 - 32 ( Tiles 041 - 048 ) have large standard deviations (error bars) due to poor receiver reboots which occurred on the 25th and 10th December 2019 respectively.}
    \label{fig_mean_and_stddev_201912_xy}
\end{figure*}

The near real-time pipeline reduces daily calibrator observations, fits their phases and amplitudes with first order polynomials (Fig.~\ref{fig_fit_example}) and inserts the resulting cable delays and intercepts into the CALDB. These fitted cable delays can be plotted as a function of time to monitor the long-term stability of the instrument. If the system is stable, the slope should be approximately constant over long periods of time (timescales of weeks or even months). Figures~\ref{fig_mwa_month_stability_extended} and~\ref{fig_mwa_month_stability_compact} show fitted delays (in nanoseconds) as a function of time for selected sixteen MWA tiles in the extended and compact configurations respectively. It can be seen that the instrument remains very stable over many weeks. A compilation of such plots for all tiles is shown in Figures~\ref{fig_mean_and_stddev_201903_xy} and \ref{fig_mean_and_stddev_201912_xy} where standard deviation of fitted delays is plotted against the antenna index enabling aggregation of the system stability in a single plot.

Routine monitoring of these plots enabled the identification of problems which can remain undetected in real time plots showing power spectra of all the MWA tiles. In particular it enables the monitoring of clock signals connected to the MWA receivers and in a few cases it identified the ``drift'' of a receiver clock due to a failure at the initialisation process, which was fixed by rebooting the receiver. It was also noticed that the phases of the calibration solutions can abruptly change when an MWA receiver is power cycled and the clock latches with an accuracy of 10\,ns resulting in a step-like change of slope (corresponding to less then 3\,metres of length using speed of light in vacuum). Since this is not a large delay, with insignificant impact on data quality, cable delays below 13\,ns (4\,metres) typically remain un-corrected, which results in less than 145\,degrees of phase difference over the $30.72$\,MHz band. However, if the delay exceeds 13\,ns the cable length in the instrument setup database is updated based on the fitted value. Usually, after re-configuration between the compact and extended configurations, several tiles need cable length adjustments in the instrument description database in order to avoid large, uncorrected cable delays (fast ``phase wraps'') that, if uncorrected, reduce the MWA sensitivity. The calibration system also helped to identify situations when coaxial cables from two tiles were accidentally swapped at a receiver input during a re-configuration between the compact and extended MWA configurations causing large delays (due to cable length from a different tile being used to correct the phase). 

The MWA instantaneous observing bandwidth is $30.72$\,MHz, which is typically placed between 50 and 350 MHz.
Thus, we could not perform the fit of a straight line over the full band (starting from zero frequency). Moreover, the combination of sky and beam models used in the calibration are usually not a perfect representation of the sky and instrument. Therefore, we allowed the intercept to be a free parameter of the fit. We verified that the fitted values of the intercept are also very stable over time (excluding times when receivers are rebooted) and they are often close to zero or a multipliety of 360\degree. Hence, with the future improvements in the sky model based on the recent extensions of the GaLactic Extragalactic All-sky MWA (GLEAM) catalogue \citep{2019PASA...36...47H} we will consider constraining the intercept value to be either zero or a multipliety of 360\degree.


\subsection{Providing calibration solutions to the new MWA correlator}
\label{subsec:calsol_for_new_correlator}

The near real-time pipeline will be used to provide calibration solutions for the new MWA ``fringe-stopping'' correlator, which is currently in development (Morrison et al. in preparation). The MWA telescope is very stable (Section~\ref{subsec:monitoring_performance}) and hence one or two sets of calibrations per 24\,hour interval should be sufficient. However, if it turns out to be insufficiently accurate, the calibration solutions for the new correlator will be updated more often. 

A dedicated calibration server will be deployed at the MRO which will enable immediate direct access to visibility files generated by the MWA correlator. This will significantly speed-up the calibration process by eliminating time required to transfer data from the MRO to the PSC archive making the pipeline a truly real-time one.

\subsection{Monitoring of calibrator field images for transients and ionospheric quality}
\label{subsec:monitoring_transients}

For the last three years of the MWA operation the near-real-time calibration pipeline produced control images of the standard MWA calibrators: Pictor A (537 images); Centaurus A (342 images); Hydra A (323 images); Hercules A (197 images); and 3C444 (214 images). There are even more archival images (before 2016) reduced when populating the MWA ASVO database with the archival data, which opens a possibility of radio-transient searches in these fields over a long time baseline  similar to those performed by \citet{2011ApJ...728L..14B} with the VLA. Roughly \sfrac{1}{3} - \sfrac{1}{2} of these images are at the MWA optimal frequency of 154.88\,MHz. We have executed the \textsc{Aegean} source finder\footnote{https://github.com/PaulHancock/Aegean} on these images in order to find transient candidates and catalogue sources to a PostgreSQL database. Analysis is on-going. We are also planning to extend the existing near real-time pipeline and look for transient candidates on a daily basis. The results of these searches will be reported in a separate publication. 
Finally, such a database populated soon after the calibrator field data are collected provides an excellent opportunity to calculate the mean offset of the sources from their nominal positions in the GLEAM catalogue \citep{gleam_nhw,2015PASA...32...25W} or other catalogues and provide early information of the given night's data quality. However, in such a case the database should be populated more densely with at least one observation every hour (or more if possible) as the ionosphere can change on timescales of hours during the night.


\section{SUMMARY}
\label{sec:summary}

 The MWA ASVO calibration component opens a new avenue for researchers worldwide to download calibrated MWA data, create sky images using standard radio-astronomy software packages and analyse these images for multiple purposes. The development of the calibration component of the MWA ASVO interface is a very important contribution to the astronomical community in Australia and beyond, providing access to the MWA data archive to every researcher without requiring a deep knowledge of the instrument. Using the recent sky models obtained from the MWA data \citep{2019PASA...36...47H,gleam_nhw} it will be possible to further improve calibration solutions in the database and consequently improve the quality of the resulting images. We expect that this endeavour will \red{facilitate greater use of the system by} researchers from outside the MWA Collaboration using MWA data. 

The development of the calibration database triggered the establishment of an automated data reduction pipeline, which has been used in near real-time for daily monitoring of the quality of the MWA calibration solutions and hence the interferometric performance of the telescope. The pipeline also produces sky images which can be used for monitoring the quality of the ionosphere and looking for transient objects on a daily basis. 

Finally, this work has been a starting point to develop a database of calibration solutions for the upcoming low-frequency component of the Square Kilometre Array telescope. Based on the MWA experience, a similar calibration database was developed to store calibration solutions from the SKA-Low prototype stations Aperture Array Verification Systems (AAVS-1 and AAVS-2) and Engineering Development Array 2  (Wayth et. al. in preparation) already deployed at the MRO. This database will be further extended in order to handle more SKA-Low stations as they will soon be built at the MRO.


\begin{acknowledgements}

\subsubsection*{People}
We would like to thank the anonymous referee for the prompt review of our manuscript.

\subsubsection*{Facilities}
This scientific work makes use of the Murchison Radio-astronomy Observatory (MRO), operated by CSIRO. We acknowledge the Wajarri Yamatji people as the traditional owners of the Observatory site. 

\subsubsection*{Funding}
Support for the operation of the MWA is provided by the Australian Government (NCRIS), under a contract to Curtin University administered by Astronomy Australia Limited. Development of the MWA ASVO was funded via the Australian Research Data Commons (ARDC), administered by Astronomy Australia Limited. Parts of this research were supported by the Australian Research Council Centre of Excellence for All Sky Astrophysics in 3 Dimensions (ASTRO 3D), through project number CE170100013. This work was further supported by resources provided by the Pawsey Supercomputing Centre with funding from the Australian Government and the Government of Western Australia. DLK was supported by NSF grant AST-1816492.

\subsubsection*{Software}
We acknowledge the work and support of the developers of the following following Python packages: Astropy  \citep{astropy:2013,astropy:2018}, Numpy \citep{2011CSE....13b..22V},  Scipy \citep{2020SciPy-NMeth}, Matplotlib \citep{Hunter:2007} and AegeanTools \citep{2018PASA...35...11H}. We acknowledge developers of the MWA\_Tools library. This research has made use of NASA's Astrophysics Data System. 

\end{acknowledgements}

\bibliographystyle{pasa-mnras}
\bibliography{caldb_mwa_asvo}

\end{document}